\documentclass[11pt,a4paper]{article}
\usepackage[T1]{fontenc}
\usepackage[utf8]{inputenc}
\usepackage{mathptmx}
\usepackage[margin=2.5cm]{geometry}
\usepackage{amsmath,amssymb}
\usepackage{graphicx}
\usepackage{booktabs}
\usepackage[font=small,labelfont=bf]{caption}
\usepackage{microtype}
\usepackage{xcolor}
\usepackage[colorlinks=true,linkcolor=blue!50!black,citecolor=blue!50!black,urlcolor=blue!50!black]{hyperref}
\usepackage{xurl}
\hypersetup{pdftitle={Monetary Policy Forecast Errors and Inflation Compensation under Sovereign Stress in Brazil},pdfauthor={Giovana Katsuki Murata; Tommy Ken Ishimatsu Goto}}

\title{Monetary Policy Forecast Errors and Inflation Compensation under Sovereign Stress in Brazil}
\author{Giovana Katsuki Murata \qquad Tommy Ken Ishimatsu Goto\\[4pt]
\small Inteli -- Instituto de Tecnologia e Lideran\c{c}a, S\~ao Paulo, Brazil\\
\small \href{mailto:giovana.murata@sou.inteli.edu.br}{giovana.murata@sou.inteli.edu.br} \quad
\href{mailto:tommy.goto@sou.inteli.edu.br}{tommy.goto@sou.inteli.edu.br}}
\date{October 2026}

\begin{document}
\maketitle

\begin{abstract}
This paper examines whether inflation compensation around Brazilian monetary policy decisions is associated differently with policy forecast errors when sovereign spreads are relatively elevated. Public Tesouro Direto quotations, meeting-specific Focus forecasts and a lagged EMBI indicator yield 68 three-year and 60 five-year event observations. The outcome is a matched-horizon retail yield proxy; the policy variable is the announced Selic target minus a preceding-week survey mean. At three years, a 10-basis-point forecast error is associated with 4.14 basis points less compensation repricing in the high-spread state than in the low-spread state, but the restricted wild cluster test gives a two-sided $p$-value of 0.1018. The estimated total high-state response is $-0.42$ basis points, with a pointwise 95\% interval of $[-2.25,\ 1.41]$. The five-year interaction is sensitive to pandemic meetings and changes sign when 2020 is excluded. A common sample of 43 events does not establish a maturity gradient. A supplementary comparison on extended windows is also sensitive to the policy measure. The evidence is therefore inconclusive about state dependence. Survey timing, sparse maturity support and daily retail quotations prevent these associations from identifying a structural monetary shock, a pure inflation-risk premium or fiscal dominance.

\end{abstract}

\section{Introduction}

Does inflation compensation respond differently to monetary policy news when sovereign risk is relatively elevated? In Brazil, this question connects inflation targeting to the interaction between interest rates, public debt and the exchange rate. Higher policy rates may reduce inflationary pressure through conventional channels while also increasing debt-service costs and sovereign risk. The mechanism in Blanchard (2004) motivates the possibility of an unconventional response, but does not imply that any positive response of inflation compensation identifies fiscal dominance.

This paper investigates the question with a fully executed event study based on publicly accessible data. The baseline policy variable is the error in a lagged, meeting-specific Focus forecast of the Selic target. Inflation compensation is constructed from coupon-free nominal and inflation-indexed principal-only bonds offered through Tesouro Direto. A high-spread indicator compares a strictly pre-decision EMBI observation with its trailing historical median. The resulting regressions describe how compensation changes around scheduled Copom decisions covary with forecast errors in two relative sovereign-risk states.

The empirical contribution is deliberately narrow. Goncalves et al. (2025) already examine Brazilian market-based inflation measures at one, two, three and five years, together with exchange rates and CDS, using identification through heteroskedasticity. Their main analysis is market based; Focus appears as a robustness exercise. This paper does not claim priority for using Brazilian inflation compensation or sovereign spreads. It adds an auditable public-data implementation of a lagged stress interaction, reports the total response within each state, and shows how incomplete maturity coverage and a small number of influential decisions affect the interpretation.

The three-year horizon provides the largest usable sample among the candidate integer-year maturities, with 68 observations. It was selected as primary on coverage grounds before estimating the outcome regressions. The secondary five-year sample contains 60 observations. For a 10-basis-point positive policy forecast error, the three-year high-minus-low interaction is $-4.14$ basis points of annual inflation compensation. Its restricted wild cluster $p$-value is 0.1018, and the total response in the high-spread state is $-0.42$ basis points with an interval spanning both signs. At five years, the negative full-sample interaction reverses when 2020 is omitted. The 43 meetings common to both maturities do not support a larger interaction at the longer horizon.

These findings provide no convincing evidence, in this design, that a higher-than-forecast Selic target raises inflation compensation under sovereign stress. They also do not establish a zero response, validate a particular degree of expectations anchoring, or refute a fiscal transmission mechanism. The intervals are wide, the policy variable can contain intervening news, and the outcome is a retail quotation proxy rather than an identified inflation expectation. The study is an exploratory empirical research note, not a preregistered structural test.

\section{Economic interpretation and related work}

\subsection{What a policy forecast error measures}

Separating anticipated policy changes from news is central to the event-study literature. Kuttner (2001) uses futures prices to distinguish expected and unexpected policy changes. Gürkaynak et al. (2005) distinguish news about the current policy setting from changes in the expected path. The present survey error primarily concerns the announced target for a particular meeting. It is not a pair of independently identified target and path factors.

For Brazil, Leigh and Xu (2025) use the difference between the Selic decision and the preceding-day Focus expectation to study lending-rate transmission, with the mean as their baseline and the median as a sensitivity. Here, the main forecast cutoff is earlier: the Friday of the preceding week. That choice reduces the reliance on a very recent survey observation whose historical public release time has not been authenticated. It also permits more intervening information. Consequently, the variable is described as a forecast error, rather than a clean announcement innovation.

Bolhuis et al. (2024) provide a public academic dataset of policy surprises constructed from market repricing. Their Brazilian series uses a one-year interest-rate swap and a close-to-close window covering the next trading day for after-market announcements. It can reflect news about both the current setting and the subsequent path. Section 6 compares that measure with Focus using identical events and an outcome window ending after the swap observation. The two regressors are economically different even when expressed in the same units.

Central bank information remains a competing explanation for announcement-associated movements. Nakamura and Steinsson (2018) and Jarociński and Karadi (2020) show why policy announcements may convey information about the economic outlook. Bauer and Swanson (2023) further examine predictability of measured surprises from pre-announcement information. A lagged survey and an EMBI interaction do not implement the identifying restrictions used in those studies.

\subsection{Inflation compensation and the sovereign-risk state}

Nominal and inflation-linked yields contain more than expected inflation. Research on Brazil discusses inflation expectations, risk compensation, liquidity, convexity and indexation effects (Vicente and Graminho, 2015; Araujo and Vicente, 2017). Joint term-structure models can separate components under additional assumptions (Abrahams et al., 2016); no such decomposition is estimated here. A survey--market wedge is therefore not reported as an identified inflation-risk premium.

The theoretical interaction of monetary and fiscal policies in Leeper (1991) and the Brazilian sovereign-risk mechanism in Blanchard (2004) motivate state dependence. However, an elevated sovereign spread can reflect expected losses, risk appetite, liquidity and global conditions. An EMBI indicator is not a direct observation of fiscal dominance, a debt sustainability threshold or a fiscal policy rule.

Three empirical objects must remain separate: the slope in the lower-spread state, the difference between states, and the total slope in the higher-spread state. A negative interaction can coexist with a positive total high-state slope. Even a precisely estimated sign for that total slope would not by itself reveal which structural mechanism generated it. The reporting below preserves these distinctions.

\section{Data and measurement}

\subsection{Sources, sample boundary and information dates}

The candidate calendar contains 148 scheduled Copom decisions from January 2006 through 30 July 2024. Decisions and announced targets come from the Banco Central do Brasil's official interest-rate history, including decisions that leave the target unchanged (Banco Central do Brasil, 2026b). This avoids inferring a meeting calendar solely from changes in a daily target series. The study stops at the documented end of the archived Ipeadata EMBI series, 30 July 2024 (Instituto de Pesquisa Econômica Aplicada, 2026); the last eligible decision in both main samples is 19 June 2024.

The raw Treasury snapshot contains 177,434 bond-date rows, beginning on 31 December 2004 and ending on 15 October 2026. Only observations within the study period enter estimation. Meeting-specific Focus records are obtained from the BCB's public OData endpoint, and the raw retrieval covers 2004--2024 (Banco Central do Brasil, 2026a). Downloaded payloads, source URLs, retrieval records and file checksums accompany the replication materials. The acquisition date is 20 October 2026, so this is a retrospective dataset, not a contemporaneously archived real-time panel.

The baseline outcome uses only Treasury quotations, the policy variable uses Focus and the official decision history, and the state uses Ipeadata. Supplementary controls are the US two-year Treasury yield and Brent spot price, distributed through FRED (Board of Governors of the Federal Reserve System, 2026; U.S. Energy Information Administration, 2026). PTAX and EMBI changes are auxiliary outcomes; they are not used to classify a joint structural mechanism. The market-surprise comparison uses the Brazilian series released by Bolhuis et al. (2024).

\subsection{Matched-horizon retail inflation compensation}

The Treasury data record morning purchase and sale yields and prices (Secretaria do Tesouro Nacional, 2026a; Secretaria do Tesouro Nacional, 2026b). The baseline uses the morning purchase yield for \emph{Tesouro Prefixado} (coupon-free nominal LTNs) and \emph{Tesouro IPCA+} without coupons (NTN-B Principal). Coupon-paying securities are excluded. Finite yields greater than $-100$\%, positive purchase prices and positive remaining maturities are required. Negative real yields are retained when valid.

Let ${y}_{d}^{j}\left({\tau }\right)$ denote an annual effective quoted yield in decimal units for type $j\in \{N,R\}$ on quotation date $d$. The baseline maturity is actual days to maturity divided by 365.25. Define

\begin{equation}
A_d^j(\tau) = \tau \log\left[1 + y_d^j(\tau)\right].
\label{eq:1}
\end{equation}

When $h$ coincides with an observed node, that node is used directly. For a horizon bracketed by distinct observed maturities ${\tau }_{L}<h<{\tau }_{U}$, linearly interpolate ${A}_{d}^{j}$ with weight $w=\left({h-{\tau }_{L}}\right)/\left({{\tau }_{U}-{\tau }_{L}}\right)$ and recover

\begin{equation}
\hat{y}_d^j(h) = \exp\left( \frac{(1-w)\,A_d^j(\tau_L) + w\,A_d^j(\tau_U)}{h} \right) - 1.
\label{eq:2}
\end{equation}

This is interpolation of the negative log discount factor implied by the chosen maturity convention. There is no extrapolation. An isolated node supports only its own exact maturity. The quoted-yield inflation-compensation proxy is

\begin{equation}
b_d(h) = \frac{1 + \hat{y}_d^N(h)}{1 + \hat{y}_d^R(h)} - 1.
\label{eq:3}
\end{equation}

The compounded ratio avoids equating compensation mechanically with an arithmetic yield difference. It does not, however, remove the economic and institutional differences between the two securities.

The source yields are reported to two decimal places in percent, corresponding to a one-basis-point input grid. Retail spreads, relative liquidity and inflation indexation can affect their movements. Historical settlement and indexation cash flows are not reconstructed exactly, so equation (3) is a transparent quotation-based approximation. In particular, the baseline tenor starts at the quotation date. A sensitivity uses 252 business days and ANBIMA's published national banking calendar (ANBIMA, 2026); it changes the tenor calculation without resolving historical settlement or indexation conventions.

\subsection{Observation windows and coverage}

For decision date $t$, the main change is

\begin{equation}
\Delta b_{h,t} = 10^4 \left[ b_{d^+(t)}(h) - b_t(h) \right],
\label{eq:4}
\end{equation}

where ${d}^{+}\left({t}\right)$ is the next observed Treasury quotation date and ${d}^{+}\left({t}\right)-t$ is at most three calendar days. Both curves must support $h$ on both exact dates. Missing curves are neither filled forward nor recorded as zero changes. The comparison is between the decision-date morning and a subsequent morning. Exact historical quote timestamps are unavailable, so it includes more news than the policy announcement alone.

Table 1 reports the usable events. The three-year horizon has the greatest coverage and is the primary outcome. Five years is secondary; one, two and four years are exploratory. Seven- and ten-year horizons have no eligible paired support under this construction and are not estimated. The sample starts at different dates across maturities, and coverage has gaps within the reported ranges. Figure 1 makes those gaps visible.

\begin{table}[!tbp]
\centering
\caption{Observed event support by horizon}
\label{tab:support}
\begin{tabular}{lcccc}
\toprule
Maturity & $N$ & First decision & Last decision & $G$ \\
\midrule
1 year & 24 & 2014-05-28 & 2024-06-19 & 6 \\
2 years & 47 & 2013-05-29 & 2024-06-19 & 9 \\
3 years & 68 & 2012-05-30 & 2024-06-19 & 12 \\
4 years & 63 & 2015-04-29 & 2024-06-19 & 10 \\
5 years & 60 & 2015-04-29 & 2024-06-19 & 9 \\
3/5 years, common & 43 & 2015-04-29 & 2024-06-19 & 8 \\
\bottomrule
\end{tabular}
\par\medskip
\begin{minipage}{0.92\linewidth}\footnotesize
\emph{Notes:} $N$ denotes decisions and $G$ distinct observed calendar years. Candidate universe: 148 scheduled decisions during 2006--30 July 2024. Eligibility requires supported nominal and real curves on both observed quotation dates, the specified window, a valid lagged Focus vintage and a strictly pre-event EMBI state. Counts use ACT/365.25 and log-discount interpolation. Dates delimit observed support, not continuous coverage.
\end{minipage}
\end{table}

\begin{figure}[!tbp]
\centering
\includegraphics[width=0.95\linewidth]{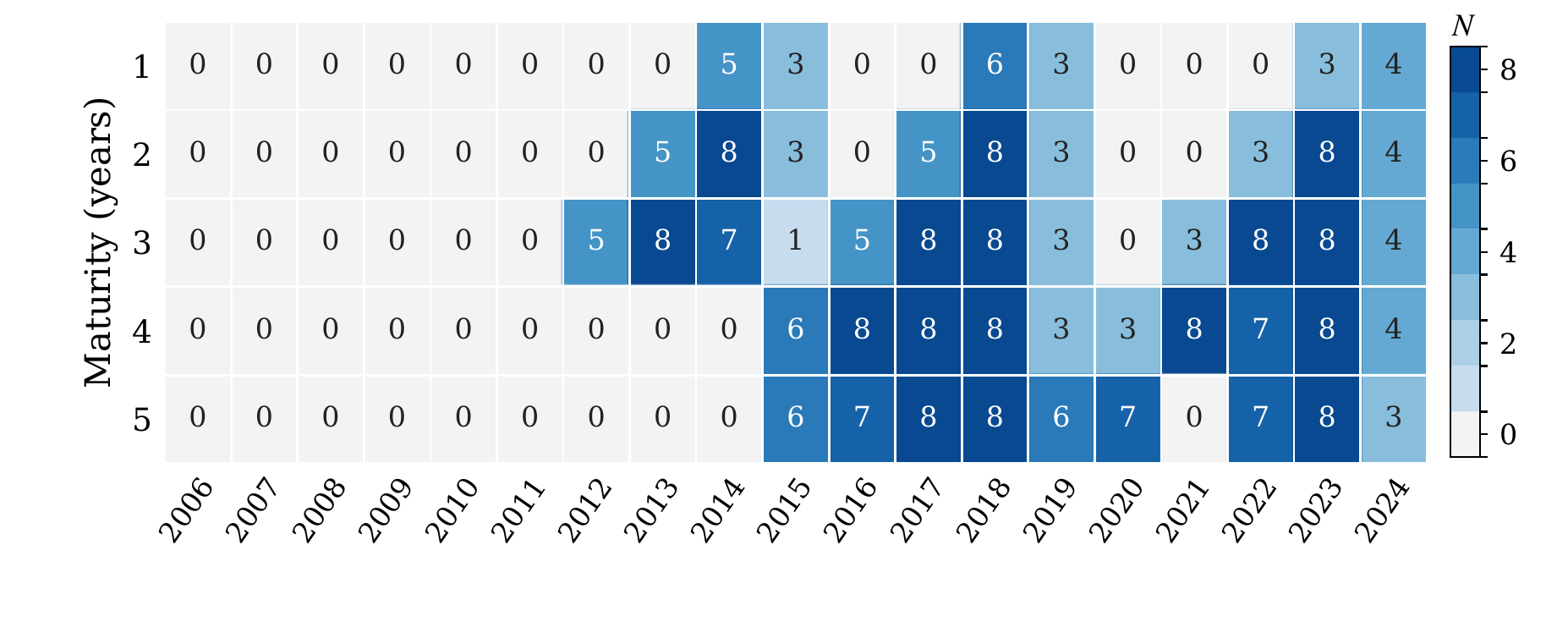}
\caption{Number of eligible decisions by year and horizon. The quotation-based samples differ across maturities and are incomplete within their date ranges.}
\label{fig:coverage}
\end{figure}

The same bracketing securities remain in place across every eligible main window. Requiring stable brackets therefore reproduces the baseline rather than supplying an independent robustness check. Maturity support still changes over time. Selection into the observable bond sample need not be random with respect to market conditions; the estimand is conditional on the available quotations.

\subsection{Policy forecast errors}

The baseline variable uses the mean Focus forecast for the specific scheduled meeting. Meeting identifiers are matched to the official within-year decision order. The source's baseline calculation group, baseCalculo=0, is retained. For each decision, select the latest survey date no later than the Friday of the preceding week, with a maximum vintage age of ten calendar days. Selected vintages in the event panel are at most six days old. If ${i}_{t}$ is the announced target and ${\overline{f}}_{v\left({t}\right),t}$ is that survey mean, both in annual percentage points, define

\begin{equation}
e_t = 100\left[ i_t - \bar{f}_{v(t),t} \right], \qquad z_t = e_t/10.
\label{eq:5}
\end{equation}

Thus ${e}_{t}$ is in basis points and coefficients on ${z}_{t}$ are per ten basis points of forecast error. The error concerns the announced target level, not the full target change. It can be nonzero at a meeting that holds the rate constant.

The BCB's historical documentation describes a weekly Monday publication based on observations through the preceding Friday (Banco Central do Brasil, 2016). That motivates the conservative date cutoff, but does not authenticate the exact public release time of every meeting-specific mean now returned by the retrospective API. Survey dates are not treated as release timestamps. Moreover, survey aggregates can retain older individual submissions. The baseline should therefore be understood as an error relative to a lagged survey vintage, not an exhaustively verified public information set available in real time.

Two sensitivities use the median from the same vintage and the mean from the last survey date before the decision. The latter is labelled a near-date survey comparison; its historical public availability is not assumed. The mean generates 47 nonzero errors in the three-year sample and 38 in the five-year sample. The corresponding median errors are nonzero at only nine and five meetings, respectively.

\subsection{A predetermined relative sovereign-risk indicator}

Let $q\left({t}\right)$ be the date of the last nonmissing EMBI observation strictly before the decision. Denote the spread on that date by ${E}_{q\left({t}\right)}$. After dropping missing observations, compute a median using the 504 observations strictly preceding $q\left({t}\right)$:

\begin{equation}
H_t = \mathbf{1}\left\{ E_{q(t)} > \operatorname{median}\left( E_{q(t)-504}, \ldots, E_{q(t)-1} \right) \right\}.
\label{eq:6}
\end{equation}

Subscripts inside the median index observations, not calendar days. The current pre-event spread is excluded from its own threshold. At least 504 earlier observations and a last-observation age of at most seven calendar days are required. This avoids using event-window outcomes to choose the state.

The classification is relative to a moving historical benchmark. An event with a lower absolute spread can be in the high state while an earlier event with a larger spread is in the low state. The terms ``high'' and ``low'' below refer to this rule. They do not assign a structural fiscal regime. The three-year sample has 42 low-state and 26 high-state events; the five-year sample has 37 and 23.

\section{Specification and inference}

Each horizon is estimated by ordinary least squares:

\begin{equation}
\Delta b_{h,t} = \alpha_h + \beta_h z_t + \gamma_h H_t + \delta_h (z_t H_t) + u_{h,t}.
\label{eq:7}
\end{equation}

The low-state slope is ${\beta }_{h}$, the difference between states is ${\delta }_{h}$, and the total high-state slope is ${\beta }_{h}+{\delta }_{h}$. These are conditional linear associations. A causal interpretation would require the policy error and its state interaction to be unrelated to other determinants of repricing in the observation window, an assumption the available data cannot establish.

Reported 95\% pointwise intervals use a CR1 covariance matrix clustered by calendar year and a $t(G-1)$ critical value, where $G$ is the number of observed years. Year clustering permits dependence among meetings within a year but assumes independent cluster contributions across years. There are only 12 years in the three-year baseline and nine in the five-year baseline. Inference for the sum ${\beta }_{h}+{\delta }_{h}$ uses the full covariance, including the covariance between the two coefficients.

For the principal comparisons, two-sided $p$-values also come from a null-imposed restricted wild cluster bootstrap with one Rademacher weight per year, 9,999 replications and a fixed recorded seed. Each contrast is tested under its own restricted null and studentized using the same clustered covariance. The reported $p$-value is $\left({1+\text{exceedances}}\right)/\left({1+B}\right)$. The approach follows the clustered-bootstrap motivation of Cameron et al. (2008), but does not eliminate the limitations of few clusters or concentrated leverage. The intervals are CR1/$t$ intervals, not inversions of the bootstrap tests; they need not give identical rejection decisions.

The primary hypothesis is ${\delta }_{3}=0$. Holm-adjusted wild $p$-values are also reported for the two baseline horizon-interaction tests. All other specifications are exploratory sensitivities. HC3 standard errors, individual-event deletion and sample restrictions assess particular vulnerabilities. Neither the clustered nor the HC3 calculation propagates uncertainty from quoted-yield interpolation, the survey vintage or the estimated state threshold.

The design record was written before the outcome regressions in this revision, after inspecting source availability and coverage. It is not an external preregistration. A later correction to the swap comparison was made after a preliminary regression run because the initial outcome ended before the final price used in the swap regressor. The corrected comparison, its motivation and the unchanged baseline are documented in Section 6 and the replication protocol.

\section{Results}

\subsection{Baseline state contrasts}

Table 2 and Figure 2 report all three contrasts. At three years, the estimated low-state slope is $+3.71$ basis points of compensation per ten basis points of forecast error. The interaction is $-4.14$, leaving a total high-state slope of $-0.42$. The interaction's pointwise interval is $[-8.55,\ 0.28]$, its wild $p$-value is 0.1018, and the Holm-adjusted value is 0.2036. Thus the primary test does not reject zero at the 10\% level. Rounding the unadjusted value to 0.10 must not be interpreted as a rejection.

\begin{table}[!tbp]
\centering
\caption{Baseline associations with inflation-compensation changes}
\label{tab:baseline}
\begin{tabular}{llrcccc}
\toprule
Maturity & Contrast & Effect & 95\% CI & Wild $p$ & $N$ & $G$ \\
\midrule
3 years & Low spread & $3.71$ & $[0.17,\ 7.25]$ & $0.1098$ & 68 & 12 \\
 & High minus low & $-4.14$ & $[-8.55,\ 0.28]$ & $0.1018$ & 68 & 12 \\
 & High spread & $-0.42$ & $[-2.25,\ 1.41]$ & $0.6722$ & 68 & 12 \\
\addlinespace
5 years & Low spread & $3.76$ & $[-1.18,\ 8.70]$ & $0.1633$ & 60 & 9 \\
 & High minus low & $-6.48$ & $[-13.38,\ 0.42]$ & $0.2620$ & 60 & 9 \\
 & High spread & $-2.72$ & $[-7.38,\ 1.94]$ & $0.7543$ & 60 & 9 \\
\bottomrule
\end{tabular}
\par\medskip
\begin{minipage}{0.92\linewidth}\footnotesize
\emph{Notes:} Units are basis points of annual compensation per 10 basis points of the preceding-week mean Focus forecast error. Low is ${\beta }_{h}$; high minus low is ${\delta }_{h}$; high is ${\beta }_{h}+{\delta }_{h}$. Intervals use year-clustered CR1 covariance and $t(G-1)$. Wild $p$-values are two-sided, null-imposed restricted bootstrap tests with 9,999 draws. The Holm-adjusted interaction $p$-values are 0.2036 and 0.2620. Intervals and bootstrap tests are different inferential procedures.
\end{minipage}
\end{table}

\begin{figure}[!tbp]
\centering
\includegraphics[width=0.85\linewidth]{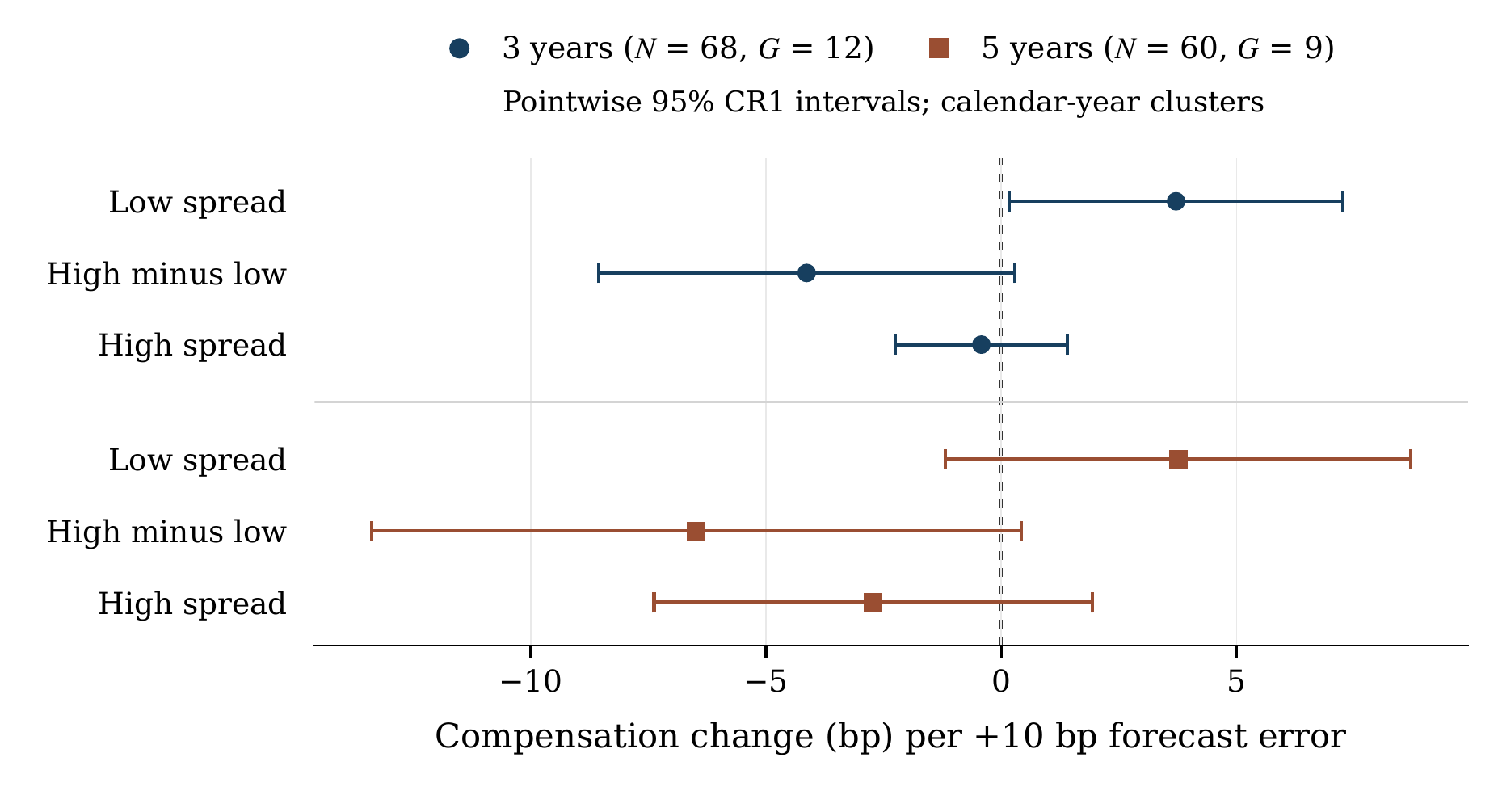}
\caption{Baseline slopes and 95\% pointwise CR1/$t$ intervals. The interaction and the total high-state slope are distinct estimands. See Table 2 for restricted wild cluster tests.}
\label{fig:slopes}
\end{figure}

The three-year total high-state interval is $[-2.25,\ 1.41]$. It includes economically different responses and does not establish that tightening lowers compensation in that state. The low-state interval excludes zero under the CR1/$t$ calculation, whereas its wild $p$-value is 0.1098. Given the small number of years, emphasis on only the more favorable inferential result would be unwarranted.

At five years, the low-state estimate is $+3.76$, the interaction is $-6.48$, and the total high-state estimate is $-2.72$. Their uncertainty is substantial: the interaction interval is $[-13.38,\ 0.42]$ with a wild $p$-value of 0.2620, and the high-state interval is $[-7.38,\ 1.94]$. The HC3 interaction $p$-value is approximately 0.565, compared with 0.097 at three years. These differences reinforce the need to examine leverage and sample composition.

\subsection{Common events and maturity comparisons}

Comparing the two full-sample interactions would mix a horizon difference with a change in the meetings observed. Restricting both outcomes to their 43 common events produces interactions of $-4.12$ at three years and $-3.04$ at five years, with wild $p$-values of 0.7172 and 0.7902. Directly regressing the five-year minus three-year change gives an interaction of $+1.07$, an interval of $[-5.11,\ 7.26]$ and a wild $p$-value of 0.7699. There is no established maturity gradient in this common sample.

Only nine common-sample events are in the high-spread state. Their mean forecast errors range from $-2.00$ to $+6.25$ basis points, and none has an absolute error of ten basis points. The ten-basis-point coefficient scale is a reporting normalization of a fitted slope; interpreting it as an observed-sized intervention in that subsample would extrapolate beyond the within-state support. The scarcity of variation helps explain the imprecision.

\subsection{Influence, survey choices and measurement checks}

Table 3 collects targeted interaction sensitivities. The most consequential change is excluding 2020 from the five-year sample: the interaction becomes $+2.28$ instead of $-6.48$. The meeting of 18 March 2020 has a $-26$ basis point forecast error, a $+28.32$ basis point outcome change, leverage of 0.525 and Cook's distance of 4.87. Omitting that event alone changes the interaction to $+2.97$. The 6 May 2020 event is also influential. Together, the two decisions account for 75.7\% of the centered sum of squared forecast errors within the five-year high state.

These are observed decisions rather than coding mistakes, and the baseline retains them. Their influence means that the estimated five-year sign is not stable evidence about an ordinary high-spread meeting. At three years, deleting one event at a time preserves a negative interaction, ranging from $-5.01$ to $-2.82$, although the total high-state slope changes sign when 27 October 2021 is omitted. Excluding 2020 leaves the three-year sample unchanged because it already contains no eligible 2020 events.

\begin{table}[!tbp]
\centering
\caption{Selected sensitivities: high-minus-low interaction}
\label{tab:sensitivities}
\begin{tabular}{lcrcccc}
\toprule
Specification & Years & Interaction & 95\% CI & Wild $p$ & $N$ & $G$ \\
\midrule
Common 3/5-year sample & 3 & $-4.12$ & $[-13.99,\ 5.75]$ & $0.7172$ & 43 & 8 \\
 & 5 & $-3.04$ & $[-13.57,\ 7.48]$ & $0.7902$ & 43 & 8 \\
\addlinespace
Focus median & 3 & $-1.02$ & $[-4.95,\ 2.91]$ & --- & 68 & 12 \\
 & 5 & $-4.74$ & $[-8.33,\ -1.16]$ & --- & 60 & 9 \\
\addlinespace
Excluding 2020 & 3 & $-4.14$ & $[-8.55,\ 0.28]$ & --- & 68 & 12 \\
 & 5 & $2.28$ & $[-7.53,\ 12.09]$ & --- & 53 & 8 \\
\addlinespace
Global controls & 3 & $-4.08$ & $[-8.63,\ 0.46]$ & --- & 67 & 12 \\
 & 5 & $-0.34$ & $[-8.03,\ 7.35]$ & --- & 59 & 9 \\
\addlinespace
Pre-event placebo & 3 & $1.91$ & $[-1.58,\ 5.40]$ & $0.3206$ & 68 & 12 \\
 & 5 & $-5.29$ & $[-9.47,\ -1.11]$ & $0.2657$ & 59 & 9 \\
\addlinespace
Linear yields & 3 & $-2.90$ & $[-7.81,\ 2.02]$ & --- & 68 & 12 \\
 & 5 & $-5.52$ & $[-13.02,\ 1.98]$ & --- & 60 & 9 \\
\addlinespace
DU/252 tenor & 3 & $-4.14$ & $[-8.56,\ 0.28]$ & --- & 68 & 12 \\
 & 5 & $-6.45$ & $[-13.32,\ 0.43]$ & --- & 60 & 9 \\
\bottomrule
\end{tabular}
\par\medskip
\begin{minipage}{0.92\linewidth}\footnotesize
\emph{Notes:} The outcome and coefficient units match Table 2, except that the placebo uses the previous observed morning to the decision-date morning. Intervals are year-clustered CR1/$t$. Wild tests are reported only for comparisons for which they were computed; a dash is not a zero or a nonsignificant result. Each specification changes the named feature. Common events number 43 at both horizons. The median has only nine and five nonzero forecast errors. Excluding 2020 changes no three-year observations.
\end{minipage}
\end{table}

Using the survey median gives interactions of $-1.02$ and $-4.74$, but the large loss of nonzero policy errors prevents treating that exercise as strong independent confirmation. Using the last pre-decision survey mean gives $-4.30$ and $-6.67$; this near-date comparison retains the release-time qualification. Requiring the post-decision quotation to be exactly the next calendar day gives $-4.66$ and $-6.65$ on 64 and 56 events.

Global controls are the change in the US two-year yield and the log change in Brent, from the preceding weekday close to the decision-date close. These are daily sensitivities, not verified common-clock controls. With them, the interaction is $-4.08$ at three years and $-0.34$ at five years. The five-year change is not explained solely by the loss of one observation: dropping that observation without adding controls leaves an interaction near $-6.25$.

Linear interpolation of effective yields, rather than their implied log discount factors, gives interactions of $-2.90$ and $-5.52$. Changing the tenor convention to 252 business days gives $-4.14$ and $-6.45$. The latter changes individual main-window outcomes by less than 0.18 basis points at either reported horizon. This limited sensitivity concerns the tenor rule only; it does not validate the retail proxy as a wholesale zero-coupon breakeven or remove indexation effects.

The pre-event placebo uses compensation from the previous observed morning to the decision-date morning. Its interaction is $+1.91$ at three years and $-5.29$ at five years, with wild $p$-values of 0.3206 and 0.2657. The five-year placebo is of similar magnitude to the main estimate. Wide bootstrap uncertainty does not turn this into proof of no anticipation or no confounding. The daily design remains exposed to information arriving outside the announcement.

\section{A comparison with a published swap-surprise series}\label{sec:swap}

The Brazilian data released with Bolhuis et al. (2024) provide 110 scheduled-meeting observations from April 2009 to December 2022 in the downloaded files. This country-specific coverage is narrower than the overall panel advertised on the authors' website. The raw mps series measures one-year swap repricing in basis points; the baseline comparison does not use the orthogonalized or information-decomposed columns. Two available releases agree on the Brazilian raw surprise values and allow the original decision date to be matched to the market observation date. Zero values are preserved as supplied.

Timing requires more than matching calendar labels. For Brazilian after-market decisions, the swap change runs from the decision-date close to the following market close. The Treasury quotation on that following morning precedes the end of the regressor. The comparison therefore extends the outcome to the first observed Treasury morning \emph{strictly after} the swap's market date. The swap market date must match the first post-decision Treasury quotation date, and the total extended window must be no longer than four calendar days. This excludes inconsistent holiday matches and ensures that the outcome ends after the regressor, while admitting additional non-announcement news.

An initial exploratory run used the shorter morning window. Its swap-comparison coefficients are superseded and are not reported as evidence. The correction was prompted by the ordering of the observations, not the resulting coefficient size. The main preceding-week Focus analysis is unchanged. Within this section, Focus and the swap measure use identical extended outcomes and identical events.

\begin{table}[!tbp]
\centering
\caption{Survey and swap measures on matched extended windows}
\label{tab:swap}
\begin{tabular}{clccrcccc}
\toprule
 & & & & \multicolumn{3}{c}{High minus low} & \multicolumn{2}{c}{High spread} \\
\cmidrule(lr){5-7}\cmidrule(lr){8-9}
Years & Predictor & $N$ & $G$ & Effect & 95\% CI & Wild $p$ & Effect & Wild $p$ \\
\midrule
3 & Focus & 49 & 9 & $0.57$ & $[-6.53,\ 7.68]$ & $0.8820$ & $3.78$ & $0.1442$ \\
3 & Bolhuis et al. & 49 & 9 & $-4.36$ & $[-8.96,\ 0.25]$ & $0.0467$ & $-2.31$ & $0.1519$ \\
\addlinespace
5 & Focus & 43 & 7 & $4.06$ & $[-2.03,\ 10.15]$ & $0.2677$ & $5.00$ & $0.3702$ \\
5 & Bolhuis et al. & 43 & 7 & $-3.81$ & $[-8.20,\ 0.58]$ & $0.1843$ & $-1.80$ & $0.3990$ \\
\bottomrule
\end{tabular}
\par\medskip
\begin{minipage}{0.92\linewidth}\footnotesize
\emph{Notes:} All coefficients are basis points of annual compensation per 10 basis points of the named policy measure. Within a horizon, the two rows use identical events and an outcome ending at the first morning strictly after the swap market date. There are 49 events and nine year clusters at three years; 43 events and seven clusters at five years. Intervals are CR1/$t$, and wild $p$-values use 9,999 restricted draws. These are exploratory comparisons, without a multiplicity adjustment across all supplementary tests.
\end{minipage}
\end{table}

At three years, the swap-based interaction is $-4.36$ with a wild $p$-value of 0.0467, while the Focus interaction on the same 49 events and longer outcome is $+0.57$ with a $p$-value of 0.8820. The swap interaction's CR1/$t$ interval is $[-8.96,\ 0.25]$, illustrating again that it is not the inversion of the wild test. The total swap-based high-state slope is $-2.31$, with a wild $p$-value of 0.1519. A nominally small exploratory interaction $p$-value therefore neither establishes a negative total high-state response nor confirms the principal survey result.

At five years, the swap interaction is $-3.81$ and the Focus interaction is $+4.06$ on the same 43 extended-window events. Their wild $p$-values are 0.1843 and 0.2677. Only seven year clusters are available. The difference in significance across measures is not a test of equality between their coefficients. Their information content also differs: one is an error about the current decision relative to an older survey, while the other is repricing of a one-year market rate. Both the sample and window differ from the main specification; their separate effects have not been isolated. The results differ across policy measures on the same extended windows, without resolving structural identification.

\section{Limitations}

Four limitations determine the scope of the findings. First, data availability selects a changing subset of meetings and maturities. The three-year and five-year baselines do not describe the same event population. The missing support is part of the result, especially when a few influential pandemic meetings appear at one horizon but not the other.

Second, the outcome combines expected inflation with risk, liquidity, retail quotation and indexation components. The use of coupon-free bonds simplifies maturity matching but does not create a fully reconstructed institutional zero-coupon curve. The sensitivity to a business-day tenor is a local measurement check, not a decomposition of those components.

Third, the survey error and daily quote window do not isolate announcement-only news. The conservative survey cutoff leaves several days for new information, and the retrospective source does not establish the exact real-time release status of every statistic. The swap comparison has a better documented market observation rule but uses a different interest-rate instrument and a longer outcome window.

Fourth, inference is based on few calendar-year clusters and, for some specifications, concentrated policy variation. Resampling conditional residuals does not resolve omitted information, endogenous state selection or generated-variable uncertainty. The study reports nonsignificant, fragile and conflicting results because those features constrain the economic conclusion. It does not interpret the absence of rejection as a proof of no state dependence.

\section{Conclusion}

Public Brazilian data permit a transparent but limited test of whether inflation-compensation changes around Copom decisions differ with relative sovereign stress. At the primary three-year horizon, the estimated high-minus-low association is negative, but the principal restricted wild cluster test does not reject zero. The estimated total response in the high-spread state is close to zero and imprecise. At five years, influential 2020 decisions determine the sign of the interaction. A common-event comparison does not establish a maturity gradient, and a supplementary swap-based measure gives different results from the lagged survey on matched extended windows.

The substantive conclusion is uncertainty about state dependence in this public-data design. The estimates do not support a claim that a higher-than-forecast Selic target raises inflation compensation under sovereign stress, and they do not identify or rule out fiscal dominance. The contribution is the documented empirical exercise: observed quotations, explicit support restrictions, a strictly lagged relative-risk state, correctly reported state contrasts and a reproducible account of the sensitivity to influential events and timing choices.

\section*{Data and code availability}
\addcontentsline{toc}{section}{Data and code availability}

The replication package is available at [REPLICATION PACKAGE DOI OR URL---INSERT BEFORE SUBMISSION]. It contains source retrieval instructions, the observed event panel, analysis code, SHA-256 checksums, the recorded design choices, and scripts that regenerate every table and figure. No proprietary terminal is required. Public accessibility is distinct from unrestricted redistribution: the swap-surprise data of Bolhuis et al. (2024) are made available by their authors for academic use with citation, and their original spreadsheets are not redistributed in the replication package. Provider conditions remain applicable to external datasets; the manuscript does not assign them a new blanket license.

\section*{Use of generative AI tools}
\addcontentsline{toc}{section}{Use of generative AI tools}

Generative AI tools were used to assist with code development and manuscript editing. The authors reviewed all code, results and text and take full responsibility for the content of this paper.

\appendix
\section{Replication and inferential details}

\subsection{Audit trail and reconstruction order}

Source manifests retain URLs, retrieval information and SHA-256 checksums. Reconstruction forms nominal and real curves, checks maturity support on exact quote pairs, joins lagged forecasts and states, and records each exclusion. The full within-year calendar, including rate holds, determines survey meeting keys before date restrictions. Tables and figures are generated from the estimates.

There are 36 fitted specifications in the full results file, including exploratory maturities, separate nominal and real yield changes, and auxiliary PTAX and EMBI changes. The main paper concentrates on the two designated horizons and sensitivities that materially affect their interpretation. Separate nominal and real effective-yield changes do not add exactly to equation (3), so they are not presented as an exact additive decomposition. Auxiliary assets use their own observed daily endpoints and do not identify a joint mechanism.

Ten independent numerical and alignment checks accompany the estimation code. They reconstruct annual CR1 covariance, the variance of the total high-state contrast, HC3 uncertainty, a scalar restricted bootstrap calculation, exact outcome pairs, Focus vintages and meeting mapping, the 504-observation state, missing control observations, and the corrected swap endpoints and common samples. An additional independent curve comparison found no eligibility disagreement across 1,480 ACT event/horizon/interpolation rows. These checks assess implementation, not the substantive exogeneity of the policy variable.

\subsection{Clustered covariance and linear contrasts}

For regressor matrix $X$, residuals $\hat{u}$ and year-specific blocks $\left({{X}_{g},{\hat{u}}_{g}}\right)$, the covariance estimator is

\begin{equation}
\hat{V} = \frac{G}{G-1}\,\frac{N-1}{N-K}\,(X'X)^{-1} \left[ \sum_{g=1}^{G} X_g' \hat{u}_g \hat{u}_g' X_g \right] (X'X)^{-1},
\label{eq:8}
\end{equation}

where $K$ includes the intercept. A contrast vector $c$ has estimate $c'\hat{\theta }$ and standard error $\sqrt{c'\hat{V}c}$. For the high-state slope, $c$ selects both the policy coefficient and its interaction. Omitting their covariance would give incorrect uncertainty.

For each bootstrap test, the restricted least-squares fit imposes $c'\theta =0$. Each simulated outcome adds its restricted fitted value to the restricted residual multiplied by one random sign for the entire year. The unrestricted model is then refitted, and the absolute studentized statistic is compared with the observed statistic. The seed is 20261020 plus a deterministic hash-derived offset for the named specification and contrast. The fixed draws aid reproducibility; with few years, the underlying set of possible sign patterns is limited.

\subsection{Interpretation of versioned data}

Offline reconstruction uses the archived inputs. Fresh downloads can incorporate source revisions and constitute a new data version; checksums distinguish them from the validated snapshot. The analysis end date is explicit. The protocol preserves the swap-window correction described in Section 6. All reported estimates use observed data.

\section*{References}
\addcontentsline{toc}{section}{References}
\begingroup
\small
\setlength{\parindent}{-1.5em}\setlength{\leftskip}{1.5em}\setlength{\parskip}{0.35em}
Michael Abrahams, Tobias Adrian, Richard K. Crump, Emanuel Moench, and Rui Yu. Decomposing real and nominal yield curves. \emph{Journal of Monetary Economics}, 84: 182--200, 2016. DOI: 10.1016/j.jmoneco.2016.10.006. URL \url{https://doi.org/10.1016/j.jmoneco.2016.10.006}.

ANBIMA. Feriados nacionais, 2026. URL \url{https://www.anbima.com.br/feriados/feriados.asp}. Published national banking calendar, 2001--2099; retrieved 20 October 2026.

Gustavo Silva Araujo and José Valentim Machado Vicente. Estimação da inflação implícita de curto prazo. Trabalhos para Discussão 460, Banco Central do Brasil, 2017. URL \url{https://www.bcb.gov.br/pec/wps/port/TD460.pdf}.

Banco Central do Brasil. Expectativas de mercado. Série Perguntas Mais Frequentes 13, Banco Central do Brasil, 2016. URL \url{https://www.bcb.gov.br/content/controleinflacao/Documents/Expectativas\_mercado/FAQ13-Expectativas-de-Mercado.pdf}.

Banco Central do Brasil. Sistema de expectativas de mercado: ExpectativasMercadoSelic, 2026a. URL \url{https://olinda.bcb.gov.br/olinda/servico/Expectativas/versao/v1/odata/ExpectativasMercadoSelic}. Public OData endpoint; 2004--2024 survey-date records retrieved 20 October 2026.

Banco Central do Brasil. Histórico das taxas de juros: decisões do Copom, 2026b. URL \url{https://www.bcb.gov.br/api/servico/sitebcb/historicotaxasjuros}. Official historical decision data; retrieved 20 October 2026.

Michael D. Bauer and Eric T. Swanson. A reassessment of monetary policy surprises and high-frequency identification. \emph{NBER Macroeconomics Annual}, 37: 87--155, 2023. DOI: 10.1086/723574. URL \url{https://doi.org/10.1086/723574}.

Olivier Blanchard. Fiscal dominance and inflation targeting: Lessons from Brazil. NBER Working Paper 10389, National Bureau of Economic Research, 2004. URL \url{https://www.nber.org/papers/w10389}.

Board of Governors of the Federal Reserve System. Market yield on U.S. Treasury securities at 2-year constant maturity, quoted on an investment basis, 2026. URL \url{https://fred.stlouisfed.org/series/DGS2}. DGS2, distributed by the Federal Reserve Bank of St. Louis; retrieved 20 October 2026.

Marijn A. Bolhuis, Sonali Das, and Bella Yao. A new dataset of high-frequency monetary policy shocks. IMF Working Paper 24/224, International Monetary Fund, 2024. URL \url{https://www.imf.org/en/publications/wp/issues/2024/10/11/a-new-dataset-of-high-frequency-monetary-policy-shocks-556255}.

A. Colin Cameron, Jonah B. Gelbach, and Douglas L. Miller. Bootstrap-based improvements for inference with clustered errors. \emph{The Review of Economics and Statistics}, 90 (3): 414--427, 2008. DOI: 10.1162/rest.90.3.414. URL \url{https://doi.org/10.1162/rest.90.3.414}.

Carlos Goncalves, Mauro Rodrigues, and Fernando Genta. Monetary policy and inflation expectations: High-frequency evidence from Brazil. IMF Working Papers 2025/048, International Monetary Fund, 2025. URL \url{https://www.imf.org/en/publications/wp/issues/2025/02/28/monetary-policy-and-inflation-expectations-high-frequency-evidence-from-brazil-562685}.

Refet S. Gürkaynak, Brian Sack, and Eric T. Swanson. Do actions speak louder than words? The response of asset prices to monetary policy actions and statements. \emph{International Journal of Central Banking}, 1 (1): 55--93, 2005. URL \url{https://www.ijcb.org/journal/ijcb05q2a2.htm}.

Instituto de Pesquisa Econômica Aplicada. Risco país---EMBI+ Brasil, 2026. URL \url{https://www.ipeadata.gov.br/ExibeSerie.aspx?module=M\&serid=40940}. Ipeadata series JPM366\_EMBI366; underlying source J.P. Morgan; observations end 30 July 2024; retrieved 20 October 2026.

Marek Jarociński and Peter Karadi. Deconstructing monetary policy surprises---the role of information shocks. \emph{American Economic Journal: Macroeconomics}, 12 (2): 1--43, 2020. DOI: 10.1257/mac.20180090. URL \url{https://www.aeaweb.org/articles?id=10.1257/mac.20180090}.

Kenneth N. Kuttner. Monetary policy surprises and interest rates: Evidence from the Fed funds futures market. \emph{Journal of Monetary Economics}, 47 (3): 523--544, 2001. DOI: 10.1016/S0304-3932(01)00055-1. URL \url{https://www.sciencedirect.com/science/article/pii/S0304393201000551}.

Eric M. Leeper. Equilibria under `active' and `passive' monetary and fiscal policies. \emph{Journal of Monetary Economics}, 27 (1): 129--147, 1991. DOI: 10.1016/0304-3932(91)90007-B. URL \url{https://www.sciencedirect.com/science/article/pii/030439329190007B}.

Daniel Leigh and Rui Xu. Monetary policy transmission to lending rates: Evidence from Brazil. IMF Working Paper 25/152, International Monetary Fund, 2025. URL \url{https://www.imf.org/en/publications/wp/issues/2025/07/25/monetary-policy-transmission-to-lending-rates-evidence-from-brazil-568783}.

Emi Nakamura and Jón Steinsson. High-frequency identification of monetary non-neutrality: The information effect. \emph{The Quarterly Journal of Economics}, 133 (3): 1283--1330, 2018. DOI: 10.1093/qje/qjy004. URL \url{https://academic.oup.com/qje/article-abstract/133/3/1283/4828341}.

Secretaria do Tesouro Nacional. Taxas dos títulos ofertados pelo Tesouro Direto, 2026a. URL \url{https://www.tesourotransparente.gov.br/ckan/dataset/taxas-dos-titulos-ofertados-pelo-tesouro-direto}. Daily retail quotation dataset; retrieved 20 October 2026.

Secretaria do Tesouro Nacional. Taxa: dicionário de dados das taxas e preços do Tesouro Direto, 2026b. URL \url{https://tesourotransparente.gov.br/ckan/dataset/df56aa42-484a-4a59-8184-7676580c81e3/resource/1a8eb2e3-4902-4a38-a1eb-6410f23d90de/download/Taxa.pdf}. Official field definitions; consulted 20 October 2026.

U.S. Energy Information Administration. Crude oil prices: Brent---Europe, 2026. URL \url{https://fred.stlouisfed.org/series/DCOILBRENTEU}. DCOILBRENTEU, distributed by the Federal Reserve Bank of St. Louis; retrieved 20 October 2026.

José Valentim Machado Vicente and Flávia Mourão Graminho. Decompondo a inflação implícita. \emph{Revista Brasileira de Economia}, 69 (2): 263--284, 2015. DOI: 10.5935/0034-7140.20150012. URL \url{https://www.scielo.br/j/rbe/a/xVRYBjcr6sWgj8yRtcjCDBt/}.
\par\endgroup

\end{document}